# System Software: Concepts and Approach

[1]Pramod Saxena, [2]Dr. Manju Kaushik
[1]M.tech (Software Engineering), [2]Associate Professor
JECRC, University, Jaipur, India
[1]saxena.pramod36@gmail.com , [2]manju.kaushik@jecrcu.edu.in

*Abstract*— **In software industry a large number of projects continue to fail due to non technical issue such as communication gap, requirements and poor executive. The authors identify the reasons for which are available for software development life cycles fall short of dealing with them. They also proposed the system development for software development life cycle. In this paper, the concept of system development, SDLC is further explored and a number of concepts are discussed in this regard.**

## I. INTRODUCTION

It could be said that the research area of development life cycles is indeed mature. Since the early days of software engineering, this area has seen the development of a number of models and methodologies ranging from the generic waterfall
Model [3] to the more recent agile techniques [4][5].

Different approaches function to varying degrees of success depending on the scenario at hand. However, given that ICT projects persistently continue to be late and even of insufficient quality [1], one is compelled to consider the possibility that the software engineering community may have taken a wrong turn
at some point.

In a preceding paper [6], the authors of this paper pointed out a number of problems which negatively affect ICT projects. It was argued that these problems are not technical in nature but are related more to the way information is created, manipulated and used within an organization. These problems included cognitive overload, information anxiety, social tension, duplicated work, and employee burnout amongst others.

The authors proposed the formalization of the concept of knowledge contexts which are discussed in section 2 and argued that a development life cycle which focused on the maintenance of a healthy knowledge context would solve the problems mentioned above. The development of an information-driven SDLC was proposed. In such an SDLC, participants would focus mainly on the maintenance of their organization's knowledge context, intentionally relegating the
software system itself to a secondary by-product of the process.

This paper explores this concept further and identifies a number of elements which would make up an information-driven software development life cycle (ID-SDLC).

## II. KNOWLEDGE CONTEXTS

A knowledge context is defined as being the knowledge, technical or otherwise, held by any of the organizations stakeholders at a particular point in time. It is a representation of the constantly changing body of knowledge held by an entity such as an organization or knowledge context and a global knowledge context. Every stakeholder inherently has a personal knowledge context associated with him/her. Collectively, the knowledge contexts of all stakeholders in an organization form that organization's global knowledge context. This is illustrated by figure 1.

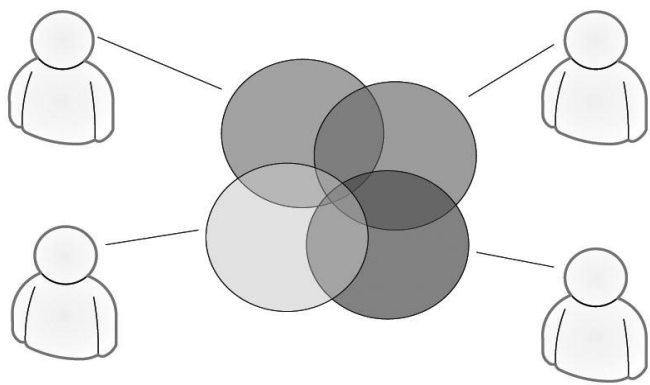

Figure 1. A graphical representation of knowledge contexts.

If one manages to formally model the concept of a knowledge context, that model could be used to manage information flow within an organization. It would become possible to give individual stakeholders the information they need, when they need it and of the level of quality which they need it. One could also reason about (or even measure) a knowledge context.

This would be useful in a number of scenarios. For Example, one could identify areas of risk whereby Certain critical knowledge such as system architecture Knowledge is only being held by one person. Such a scenario puts an organization at risk in the sense that if this person leaves, there the associated knowledge is lost.

### III. CORE ELEMENTS OF AN ID-SDLC

It is being proposed that an information driven development life cycle consist of four core elements. Namely, Information Assets, Actors, Roles, and Information Routing Groups. Other components are needed for the life cycle to come together but these four core components form the foundations on which the rest of the life cycle is being built. Information Assets are representations of knowledge. An information asset could take the form of a document, website, sound clip or any other format which would be useful in achieving one or more goals within the organization. One could easily think of a number of information assets which are used on a day-to-day basis. These include specifications, designs, test plans, source code, and so on. An information asset needs to be evaluated for its quality before being utilized. This is discussed further in section 6. Actors participate in the development of a system or have the potential of influencing it. They are capable of creating, maintaining and using information assets throughout the development life cycle. Examples of actors include developers, test engineers, managers and customers.

Roles are used as a tool of generalization within the development life cycle. It is envisaged that certain routing and evaluation of information assets can be defined through role-based rules. This would make life cycle management easier. Information Routing Groups (IRGs) were first proposed by Andrews [7] in the 1980s. An Information Routing Group (IRG) is one of a semi-infinite set of similar interlocking and overlapping groups. One group would have a number of individuals as members. These members are sometimes referred to as IRGists and would share a particular common interest. In the case of a software development organization, the interest might be a project, a programming language, a technology, and so on. Such groups are useful in an ID-SDLC because the global IRG structure would help in determining who certain information assets should be routed to for use and/or evaluation

### IV. MINI-CYCLES

The information driven approach necessitates a psychological shift from conceptualizing a life cycle as a series of activities centered around one product, to an altogether different (possibly discomforting) approach whereby the life cycle refers to the life cycle of information assets. To a certain extent, the finished product would be a bi-product of the life cycle's activities. To this effect, it is being proposed that the concept of mini-cycles be used. That is, each information asset will have one or more mini life cycles associated with it whenever particular events occur. For example, the creation of a new information asset such as a design decision would trigger a life cycle for that information asset. It would go through a number of phases and possibly trigger a number of other events within the organization. A life cycle for an information asset would typically be between an hour and a week in length depending on the amount of information contained within the asset. The aggregate of all mini-cycles would make up the whole information driven development life cycle.

When considering events related to information assets, one would typically expect such assets to be created or modified. Modification could include change in content or a change in one or more attributes of an information asset. When such events occur, it is being proposed that the information asset in question goes through a life cycle having the following four phases (also depicted in figure 2) - capture, relate, evaluate, and finally disperse.

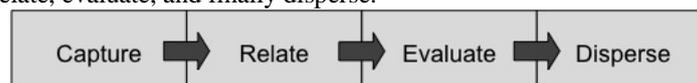

Figure 2. The envisaged stages in a mini-cycle.

The capture of an information asset entails the actual absorption of the asset into the life-cycle. This will most probably require tool support since the vast amounts of information typically utilized in a product's development is not akin to manual management. Once an information asset has been captured, it would need to be related to other information assets. For example, a design decision might override a previous decision. Such a relationship would need to be defined so that necessary actions could be taken. If this is done correctly, one can visualize the structure of interrelated information which would build up over time and the uses to which it could be put. Once relating is done, the information may need to be evaluated for quality. Quality attributes may depend on the type of content contained within the asset but a few examples include believability, consistence, objectivity, and so on. A number of possible evaluation mechanisms have been discussed but it is beyond the scope of this paper to delve into that granularity. Once the quality of the asset has been determined, it can be dispersed to actors who need it. Sticking with the example of a design decision, this may be automatically routed to developers and test engineers so that they could change their work accordingly.

### V. EVENT HIERARCHY

Events will provide the dynamic framework of the development life cycle. It is through them that mini cycles are initiated. To this end, an event hierarchy is being proposed which is split into two main categories Information Events and Stakeholder Events. The first category is concerned with events which involve information assets. Typically, this involves new information being created or existing information being changed.

Such events should invoke a mini-cycle which goes through the phases of capturing, relating, evaluating and dispersing the information asset in question (as discussed in section 4). The second category is more concerned with the upkeep of the global knowledge context when there is a change in stakeholders. This could include a new stakeholder joining the organization, an existing stake holder transferring to a new project, a stakeholder leaving the company, and so on.

In such cases, actions need to be taken in order to maintain a healthy knowledge context. Therefore in the case of a new employee, there would need to be mechanisms for getting him/her up to speed with what they need to know whilst in the case of the employee who is leaving, one would need to organize an effective Handover of knowledge so that the global knowledge Context is maintained.

Work in this area is still in preliminary stages and is likely to change. However, it is worth showing the draft event hierarchy here (figure 3). Figure 3. Initial draft of the ID-SDLC Event Hierarchy.

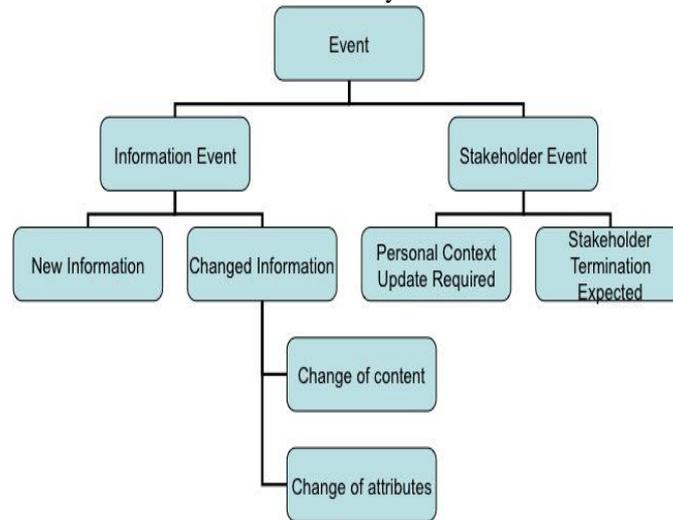

Figure 3. Initial draft of the ID-SDLC Event Hiearchy

## VI. INFORMATION QUALITY

The quality of information assets is a concept which plays a central role in the ID-SDLC. If an information asset does not exhibit a desired level of quality then a risk exists whereby, if the information assets was utilized, its inferior quality would be reflected in a final product of lower quality. Of course, one must the relevant quality attributes and the desired level of presence for each attribute. In all likelihood, quality threshold will be defined based on the information asset. For example, one would expect a design document to be consistent and correct but would other attributes such as objectivity may not apply. Mechanisms should be put in place to decide what should be done if an information asset falls below its desired quality threshold. One may decide to discard it, flag it for review, send it back to it's author, and so on. Substantial research works about into information quality have been carried out in the field of information management. One influential study by Lee et al 8 identifies fifteen key information quality attributes which are listed in table 1. It is being proposed that these quality attributes form the basis of information quality management in the IDSDLC. As stated in section 4, a number of mechanisms and ideas have been discussed in this regard but it is beyond the scope of this paper to delve into detail.

| Accessibility | Interpretability |
|---|---|
| Appropriate Amount | Objectivity |
| Believability | Relevance |
| Completeness | Reputability of Source |
| Conciseness | Security |
| Consistence | Timeliness |
| Correctness | Understandability |
| Ease of Operation | |

Table 1. Information quality attributes identified by Lee et al [8]

## VII. CONCEPT LEVELS

The concepts defined in this paper are mostly of a generic nature and work is underway to flesh them out and propose more concrete instantiations (so to speak) of the high-level concepts defined here. This will include things like information evaluation mechanisms, rule-based mechanisms for information asset dispersion, information quality metrics, product quality metrics, and so on. When this starts to take place, it is likely that different users of the life cycle would want to use a subset of these resources or may even require to have the facility to extend the life cycle for their particular situation. To this end, it is being proposed that the final life cycle offer tools and resources on four levels (figure 4).

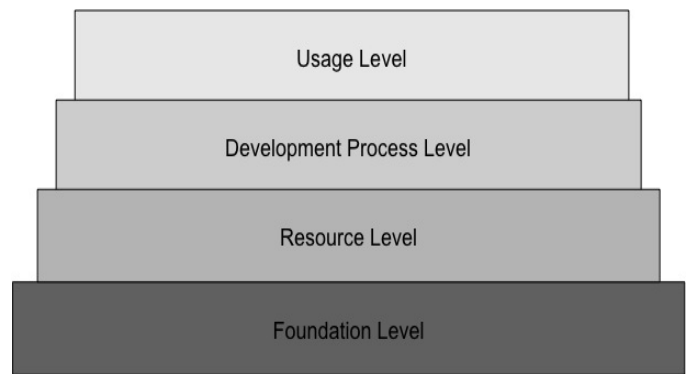

Figure 4. Proposed layered concept approach

The foundation level will include all resources which are at the core of the ID-SDLC. These will be concepts and mechanisms which cannot be customized or changed and should be used by all users of the life cycle. Examples at this level include the definition of concepts like what information assets are, what routing groups are, how these interact together, and so on. Once that is defined, one can build a resource level on top of it. As its name implies, this level will offer resources and instantiations of the concepts defined at the foundation level so as to reduce the amount of setup time one requires when using the life cycle for the first time. Take as an information asset as an example. At the foundation level, the ID-SDLC will specify what an information asset is, what information it could contain, how it can be evaluated and so on. At resource level, one would define particular information assets such as a design document, a decision, a test plan, and so on. Each of these would have specific attributes attached which will be used by any evaluation and routing mechanisms throughout a project's lifetime. At the development process level, one would be able to define mechanisms, rules and resources which are specific to the development process used within the organization. So if one is using Scrum for example, concepts like burn down charts and sprints will be defined. A number of predefined resources will be available but one could extend with new concepts or modify the existing ones. The top level is named the usage level because it will allow the user to define actual instantiations of objects for day-to-day use. This would include definition of real-life actors, roles, information assets, events, and so on.

## VIII. CONCLUSIONS AND FUTURE WORK

This paper is not intended to reach any research based conclusions but rather discuss work-in-progress in the research areas which the authors are involved in. It is believed that when fully developed, the work discussed here will be shown to contribute greatly to problems discussed in section 1.

With regards to future work in this area, it is the intention of the authors to continue to flesh out these concepts with a special focus of how they can be brought together into one coherent life cycle. This life cycle would need to have a solid basis with a number of mathematically provable properties. It should also provide metrics and measures which allow one to measure things like various properties of the knowledge context, information quality, product quality, and so on.